\documentclass[twocolumn,apj]{openjournal}

\usepackage[breaklinks,colorlinks,citecolor=blue,urlcolor=blue]{hyperref}
\usepackage{amsmath}

\usepackage[T1]{fontenc}
\usepackage{newtxtext,newtxmath}
\usepackage{graphicx}
\usepackage{amsfonts}
\usepackage{soul}
\usepackage{enumitem}
\usepackage[hyphens]{xurl}
\usepackage{multirow}
\usepackage{hyperref}

\newcommand{\orcidauthor}[3]{\author{\href{http://orcid.org/#1}{#2$^{#3}$}}}

\newcommand{\pc}{\>{\rm pc}}
\newcommand{\kpc}{\mbox{$\>{\rm kpc}$}} 

\newcommand{\Gyr}{\mbox{$\>{\rm Gyr}$}}

\newcommand\degrees{^\circ}
\newcommand{\avg}[1]{\mbox{$\left<{#1}\right>$}}

\newcommand{\feh}{\mbox{$\rm [Fe/H]$}}

\newcommand\gaia{{\it Gaia}}

\shorttitle{Chemical dissection of morpho-kinematic lopsidedness}
\shortauthors{Soumavo Ghosh et al.}

\begin{document}


\title{Chemical dissection of merger-induced $m=1$ lopsidedness in Milky Way-like galaxies\vspace{-1.5 cm}}

\orcidauthor{0000-0002-6549-7455}{Soumavo Ghosh}{1,^\dagger} 
\orcidauthor{}{Paola Di Matteo}{2}
\orcidauthor{}{Chanda J. Jog}{3}
\orcidauthor{0000-0000-0000-0000}{Neige Frankel}{4} 

\affiliation{$^{1}${Department of Astronomy, Astrophysics and Space Engineering, Indian Institute of Technology Indore, India - 453552}} 
\affiliation{$^{2}${LIRA, Observatoire de Paris, Université PSL, Sorbonne Université, Université Paris Cité, CY Cergy Paris Université, CNRS, 92190 Meudon, France}} 
\affiliation{$^{3}${Department of Physics, Indian Institute of Science, Bangalore 560012, India}}
\affiliation{$^{4}${Canadian Institute for Theoretical Astrophysics, University of Toronto, 60 St. George Street, Toronto, ON M5S 3H8, Canada}}

\thanks{$^\dagger$ Corresponding author: \href{mailto:soumavo@iiti.ac.in}{soumavo@iiti.ac.in}}

\begin{abstract} 
The Milky Way harbours a prominent $m=1$ lopsided distortion in both stellar and neutral gas distributions. On the other hand, chemo-dynamical studies have been proven to be effective in grasping the overall evolution of galaxies.
  Here, we investigate systematically the excitation and evolution of a merger-driven $m=1$ lopsidedness in a Milky Way (MW)-like host galaxy, as a function of chemical distribution of stars. Using seven  dissipationless, high-resolution $N$-body simulations of minor mergers (between a MW-like host and a satellite) under varying orbital configurations (prograde/retrograde and different orientation of the satellite orbital plane), we first show that a tidal interaction excites a prominent $m=1$ lopsidedness in the stellar density and velocity distribution of the MW-like host. Assigning, a posteriori, metallicities to stellar particles of the MW-like host based on the current observational constraints, we sub-divide the stars into metal-rich ($\feh >0$), metal-intermediate ($-0.5 \leq \feh <0$), and metal-poor ($- 0.5 < \feh $) populations. We demonstrate that metal-rich population always show a much stronger $m=1$ lopsidedness in both density and velocity distributions when compared to other two populations. This trend holds true for all minor merger model considered here, regardless of their orbital configurations. Furthermore, minor merger also triggers a transient off-centred stellar disc-dark matter halo configuration, with metal-rich population showing the highest degree of disc-halo offset. We show that the metal-rich population  which is kinematically colder (i.e. lower velocity dispersion) by construction, is more susceptible to external perturbations. Lastly, using a catalogue of photometry and metalicity for the LMC,  we show that the strength of the $m=1$ distortion (predominantly in the form of an one-arm spiral) in stars increases with metallicity as well.

\end{abstract}

\section{Introduction}
\label{sec:Intro}

The $m=1$ feature, or lopsidedness, a distortion along the azimuthal direction, is ubiquitous in the spatial distribution of stars \citep[e.g. see][]{Blocketal1994,RixandZaritsky1995,Bournaudetal2005,Reichardetal2008,Zaritskyetal2013} as well as in the distribution of neutral hydrogen (H~{\sc i}) extending well beyond the stellar disc in spiral galaxies \citep[e.g. see][]{Baldwinetal1980,RichterandSancisi1994,Haynesetal1998,Matthewsetal1998,Angirasetal2006,vanEymeren2011}. The $m=1$ lopsidedness often co-exists with other non-axisymmetric perturbations, for example, an $m=2$ bar and/or spiral arms in disc galaxies \citep[e.g., see][]{RixandZaritsky1995, Bournaudetal2005, Zaritskyetal2013}. However, past observational studies investigating the correlation (if any) of the properties of the $m=1$ lopsidedness with those of $m=2$ bar or spirals have yielded conflicting results \citep[e.g. see][]{Bournaudetal2005,Zaritskyetal2013}. The Milky Way also harbours an $m=1$ lopsidedness in the stellar as well as in the (H~{\sc i}) distribution \citep{KalberlaandDedes2008,Romeroetal2019}. 
The North-South asymmetry in the H~{\sc i} scale heights observed in the Milky Way \citep[e.g. see][]{Levineetal2006} has been modelled, and the corresponding disc lopsidedness has been shown to arise due to a halo that shows a lopsided ($m=1$) as well as an $m=2$ distortion \citep[e.g. see][]{Sahaetal2009}.
Prominent signature of lopsidedness has been detected in the H~{\sc i} velocity fields of galaxies as well \citep[e.g. ][]{Swatersetal1999,Schoenmakersetal1997,vanEymerenetal2011}. Past theoretical studies have shown that a lopsided perturbation in the density distribution can result in a kinematic lopsided feature \citep{Jog1997,Jog2002} which has later been confirmed observationally for a sample of galaxies from the WHISP (Westerbork H~{\sc i} Survey of Spiral and Irregular Galaxies) survey \citep[see][]{vanEymerenetal2011,vanEymeren2011} as well as from numerical simulations \citep[e.g.][]{Ghoshetal2022}.
\par
In the past, several physical mechanisms have been proposed for the excitation of an $m=1$ lopsidedness in galactic discs. These include  the disc response to a distorted halo \citep[]{Jog1997,Schoenmakersetal1997}, a
tidal interaction or merging of a satellite galaxy \citep[e.g. see][]{ZaritskyandRix1997,Bournaudetal2005,Mapellietal2008,Ghoshetal2022}, asymmetric gas accretion \citep{Bournaudetal2005} or an off-set disc in a spherical dark matter (hereafter DM) halo \citep{Noordermeer2001,PrasadandJog2017}. Regardless of the generating mechanism involved, the $m=1$ lopsidedness plays a pivotal role in secular evolution of disc galaxies, by facilitating the outward angular momentum transport \citep{SahaandJog2014}, thereby allowing the inflow of gas from the outer regions of a galaxy. In addition, measuring the pattern speed of the lopsided asymmetry bears crucial insight about the dynamical role of the lopsidedness in secular evolution as well as in determining the generating mechanisms of the lopsidedness  \citep[e.g. see discussion in][]{Jog2011}. Using a suite of 1:10 minor merger models, \citet{Ghoshetal2022} demonstrated that the $m=1$ lopsidedness rotates at a much lower pattern speed when compared to that for the $m=2$ bar, and is in retrograde with respect to the bar pattern speed. For a detailed exposition on this topic, the reader is referred to \citet{JogandCombes2009}.
\par
Chemical cartography, by combining the chemical composition with the 6-D phase-space information of stars in the MW, has been shown to play a crucial role in comprehending the overall evolution of the MW, possibly shaped by internal and external perturbations \citep[e.g. see][]{Haydenetal2015,GaiaCollab2023}. Furthermore, a recent study by \citet{Frankeletal2025}, using the \gaia\ XP spectra, hinted toward a presence of metallicity-dependent $m=1$ lopsided distortion in the stellar density distribution of the LMC. Whether this is a genuine trend or just a visual bias -- it is yet to be verified quantitatively. The metal-poor stars tend to be older and dynamically hot, that is, having higher velocity dispersion \citep[e.g. see][]{Haywoodetal2013}. In addition, recent numerical simulations have shown that an $m=2$ bar and the associated boxy/peanut bulge tend to be stronger in younger and dynamically-cold stellar population \citep[e.g. see][]{Ghosetal2023,Ghoshetal2024}. However, any such systematic study, investigating the variation of the properties of the $m=1$ lopsidedness as a function of metallicity of stars is still missing and is certainly worth investigating. We pursue it in this work.
\par
Here, we carry out a detailed study of the metallicity variation of the merger-driven $m=1$ lopsidedness in density and velocity fields of stars in MW-like host galaxies. For this work, we make use of a suite of minor merger models (of mass ratio 1:10) with varying orbital configurations where an $m=1$ lopsidedness is triggered in the MW-like host by the tidal interaction with the satellite. In particular, we focus on the dynamical response of the stellar disc of the host galaxy to such tidal interactions, and quantify the $m=1$ lopsidedness in stellar density and velocity fields for different stellar populations with varying metallicity bins as the models evolve with time. In addition, we investigate the excitation of an off-set between the baryonic disc (of varying metallicity) and the DM halo, and follow their dynamical linkage with the merger event.
\par
The rest of the paper is organised as follows.
Section~\ref{sec:simu_setup} provides a brief description of the minor models, their equilibrium configuration as well as the orbital configurations. Section~\ref{sec:chemical_dissection_lopsided} and sect.~\ref{sec:chemical_dissection_Kinelopsided} provide the details of the morphological and kinematic lopsidedness as a function of stellar metallicity. Section~\ref{sec:disk_halo_offset} contains the details of the disc-DM halo off-set and the associated temporal evolution. Section~\ref{sec:met_kine_connection} shows the linkage between metallicity and kinematics of the stars and the effect of kinematics on the resulting lopsidedness. Section~\ref{sec:implications_LMC} provides results pertaining to the metallicity variation of $m=1$ asymmetric structure of stars in the LMC. Section~\ref{sec:discussion} contains the discussion and sect.~\ref{sec:conclusion} summarises the main findings of this work.

\section{Minor merger models}
\label{sec:simu_setup}

For this work, we use a total of 7 dissipationless, high-resolution $N$-body simulations of a MW-type galaxy undergoing a minor merger with a satellite galaxy (of mass ratio 1:10) under varying orbital configurations. The details of the equilibrium models as well as orbital configuration are given in \citet{Pagninetal2023}. For the sake of completeness, we briefly mention the details of the minor merger models used here. Furthermore, these models have been extensively used in studying the distribution of globular clusters in kinematic spaces \citep{Pagninetal2023} and the metallicity distribution of halo stars in MW-like host galaxies as a result of mergers \citep{Morietal2024}.

\subsection{Equilibrium model}
\label{sec:equilibrium_model}

The MW-like host galaxy consists of a thin, an intermediate, and a thick stellar disc - each of them is represented by a Miyamoto-Nagai density distribution \citep{MiyamatoandNagai1975} with a total mass $M_*$, characteristic length $R_{\rm d}$ and height $h_z$. This configuration broadly mimics the Galactic thin disc, the young thick disc, and the old thick disc, respectively \citep[e.g. see][]{Haywoodetal2013,DiMatteo2016}. The three-component stellar disc is further embedded in a DM halo, modelled as a Plummer sphere \citep{Plummer1911} with a characteristic mass $M_{\rm dm}$ and length $a_{\rm dm}$. The structural parameters for each of these components are mentioned in Table.\ref{table:key_param}. A total of $10^7$ particles are used to model the thin disc whereas $6 \times 10^6$ and $4 \times 10^6$ particles are used to model the intermediate and thick discs, respectively. The DM halo is modelled with a total of $5 \times 10^6$ particles. The satellite galaxy is a re-scaled version of the MW-like host galaxy with mass and a total number of particles ten times smaller and size reduced by a factor $\sqrt{10}$ \citep[for further details, see][]{Pagninetal2023}. The final $N$-body models of the MW-like host  and of the satellite are constructed following the iterative method described in \citet{Rodionovetal2009}.

%
\begin{table*}[htb!]
\centering
\caption{Key structural parameters for the equilibrium models.}
\begin{tabular}{ccccccc}
\hline
\hline
Component  & Mass ($\times 10^{10} M_{\odot}$) & $R_{\rm d}$ (kpc) & $h_z$ (kpc) & $N_{\rm part}$ ($\times 10^6$) & $\avg{\feh}$ (dex) & $\sigma_{\avg{\feh}}$ (dex)\\
\hline
MW: thin disc & 3.73 & 4.8 & 0.25 & 10 & 0.25 &  0.15\\
MW: intermediate disc & 2.7 & 2.0 &  0.6 & 6 & -0.26 & 0.2  \\
MW: thick disc & 1.93 & 2.0 &  0.8 & 4 & -0.62 & 0.26\\
DM halo &368 & 20 &  - & 5 & - & -  \\
\hline
\end{tabular}
\newline{
\textbf{Notes:} From left to right : component name, total mass, characteristic radius (disc scale length for the disc components and Plummer radius for the DM halo), scale height, number of particle used, mean metallicity of the MDF, and dispersion of the MDF. The structural parameters are taken from \citet{Pagninetal2023} while the metallicity values are taken from \citet{Fragkoudietal2018}.}
\label{table:key_param}
\end{table*}

\subsection{Orbital configuration of merger models}
\label{sec:merger_model_details}

Once the equilibrium models for the MW-like host and the satellite galaxy are generated, the barycentre of the satellite galaxy is translated to the
point ($x_{\rm sat}$, $y_{\rm sat}$, $z_{\rm sat}$) and assigned an initial velocity ($v_{x, \rm sat}$, $v_{y, \rm sat}$, $v_{z, \rm sat}$) in order to have a parabolic orbit where the satellite is initially at a distance of $100 \kpc$ from the MW-like host. The initial position and velocities differ in each of the minor merger model, and the values are given in Table.~2 of \citet{Pagninetal2023}. The details of the subsequent temporal evolution of the separation between the MW-like host and the satellite are provided in Appendix~\ref{appen:details_merger_history}. In addition, we varied the initial orientation of the satellite orbital plane ($\Phi_{\rm orb}$) by rotating it about the $y$-axis in such a way that it covers a range of possible orientations, including prograde orbits ($\Phi_{\rm orb}= \ 0\degrees, \ 30 \degrees, \ 60 \degrees$, where  $\Phi_{\rm orb}= \ 0\degrees$ denotes a planar prograde orbit), a polar orbit ($\Phi_{\rm orb}= \ 90 \degrees$), and retrograde orbits ($\Phi_{\rm orb}= \ 120\degrees, \ 150 \degrees, \ 180 \degrees$, where  $\Phi_{\rm orb}= \ 180 \degrees$ denotes a planar retrograde orbit). This, in turn, will help us to carry out a systematic study of the response of a host galaxy to a 1:10 minor merger scenario. Furthermore,  we mention that the minor merger models used here, both the satellite and the MW-type host galaxy can react to the interaction and experience
kinematical heating, tidal effects, and dynamical friction; thereby making them more realistic \citep[for further details, see][]{Jean-Baptiste2017,Pagninetal2023}.

\par
All simulations were run by using a TreeSPH code by \citet{SemelinandCombes2002} which employs a hierarchical tree method \citep{BarnesandHut1986} with opening angle $\theta = 0.7$ to compute the gravitational forces and include terms up to the quadruple order in the multipole expansion. A Plummer potential was employed for softening the gravitational forces with a softening length $\epsilon = 50 \pc$. We adopted a fixed time step of $\Delta t = 2.5 \times 10^5$ yr to integrate the equations of motion using a leapfrog algorithm.
\par
To maintain a consistency with previous works, we followed the same scheme of nomenclature as used in \citet{Pagninetal2023} where each of the minor merger models is referred via a unique string ${\rm MWsat \_n}N\_ \Phi \beta$ where $N$ denotes the number of satellites merged with the host ($N =1$ here for all models) and $\beta$ denotes the value of the $\Phi_{\rm orb}$. The same nomenclature scheme is applied throughout this paper.

\subsection{Assigning chemical abundances}
\label{sec:chemical_abundances}

\begin{figure}
\centering
\resizebox{0.95\linewidth}{!}{\includegraphics{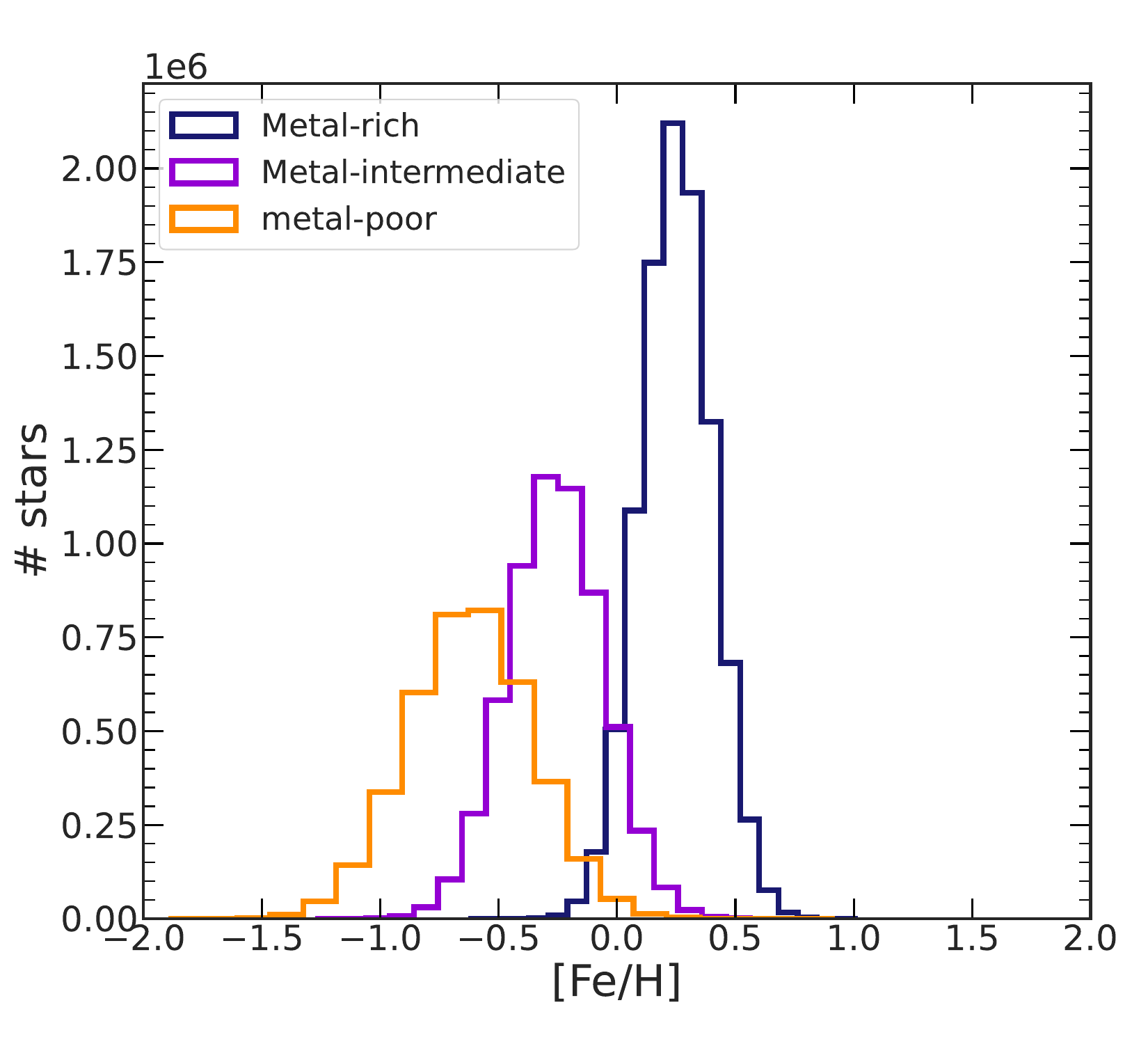}}
\caption{Metallicity distribution function (MDF) adopted for this work, at $t=0$ for the fiducial minor merger model {\rm MWsat\_n1\_$\Phi 30$}. The thin disc is associated with the metal-rich (\feh $>0$), the intermediate with the metal-intermediate ($-0.5 \leq \feh <0$), and the thick disc with the metal-poor ($\feh < -0.5$) population. For further details, see the text.}
\label{fig:mdf_allmodels}
\end{figure}

As stated in sect.~\ref{sec:Intro}, the primary goal of this work is to study the dynamical consequence of a tidal interaction (and subsequent merger) on the host galaxy as a function of metallicity of stars. However, the minor merger models analysed in this paper are dissipationless, which implies that no star formation or chemical enrichment model was included. To overcome this, we
have assigned, a posteriori, metallicities to the stellar particles of the MW-like host and the satellite based on their kinematic properties and the current observational constraints \citep[for further details, see][]{Morietal2024}. This method of assigning metallicity has been routinely followed in past works \citep[e.g. see][]{DiMatteoetal2013,Fragkoudietal2018,Khoperskovetal2018,Morietal2024,Boinetal2024}.  Following \citet{Boinetal2024}, for each of the discs (thin, intermediate, and thick), we assume a metallicity distribution function (MDF) of the form of a normal distribution having a constant (i.e. without any radial or vertical metallicity gradient within each component) mean metallicity ($\avg{\feh}$) and dispersion ($\sigma_{\avg{\feh}}$). Then, at time $t$, we assign a metallicity to a disc particle (pertaining to thin, intermediate or thick disc component) by drawing randomly from the corresponding MDF. The corresponding values of the $\avg{\feh}$ and $\sigma_{\avg{\feh}}$ for the thin, intermediate, and thick discs (of the MW-like host) are mentioned in Table.\ref{table:key_param} and are taken from \citet{Fragkoudietal2018}. Fig.~\ref{fig:mdf_allmodels} shows the metallicity distribution, generated by following the above mentioned process, for a minor merger model used in this work. Following \citet{Boinetal2024}, we associate the thin disc with metal-rich (\feh $>0$), the intermediate with metal-intermediate ($-0.5 \leq \feh <0$), and the thick disc with  metal-poor ($\rm \feh < -0.5$) populations. Although this scheme of assigning the metallicity a posteriori, as done here, is rather simplistic, nevertheless, this allows us to employ both chemical and kinematic information in studying galactic dynamics in a much more realistic fashion.

\section{Chemical dissection of the m=1 lopsidedness}
\label{sec:chemical_dissection_lopsided}

\begin{figure*}
\centering
\resizebox{\linewidth}{!}{\includegraphics{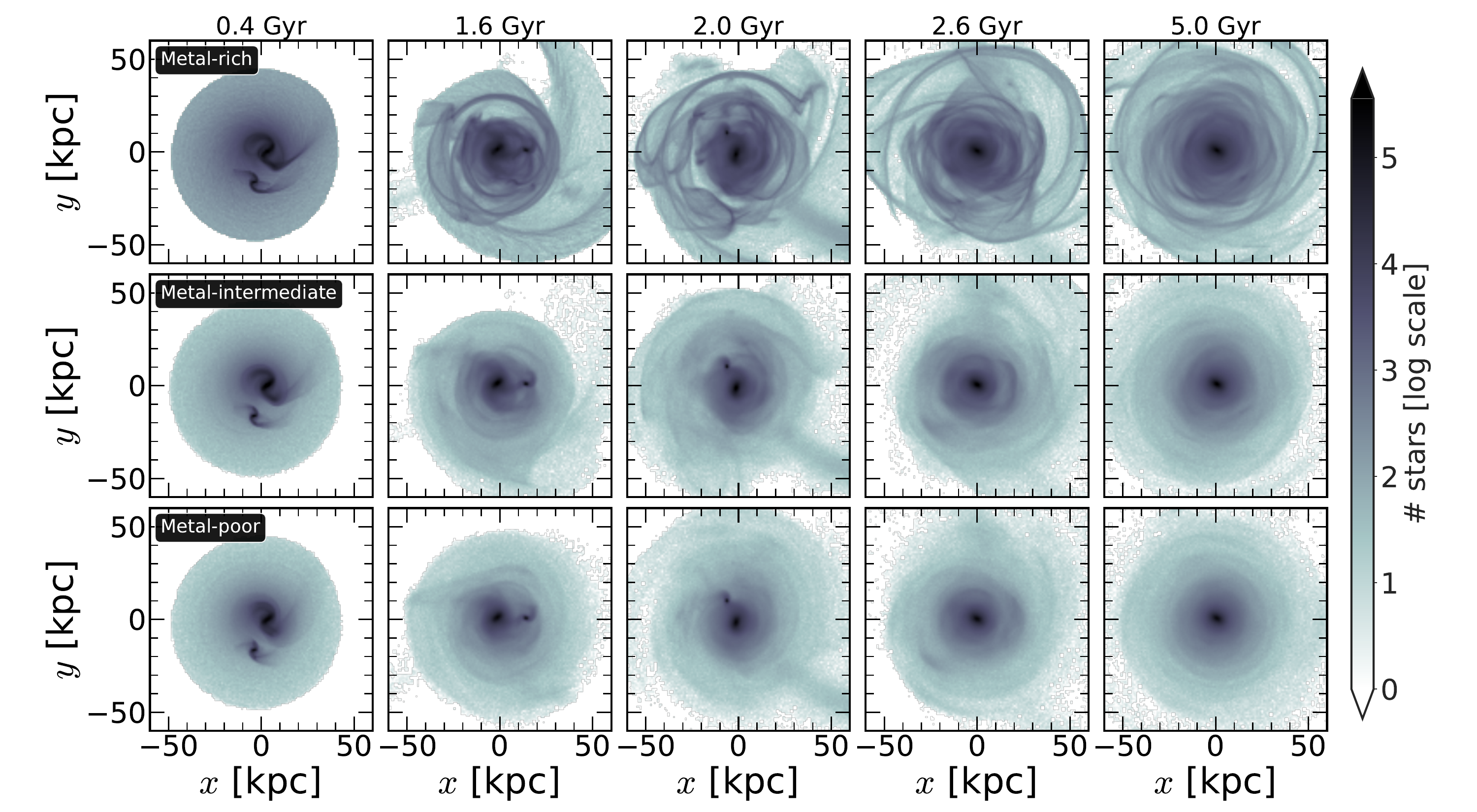}}
\caption{\textit{Chemical dissection of $m=1$ lopsidedness:} Face-on ($x-y$-plane) stellar density distribution (MW-like host+satellite), for the metal-rich (\textit{top row}), metal-intermediate (\textit{middle row}), and metal-poor  (\textit{bottom row}) populations, at different times capturing different pericentre passages and the merger epoch of the satellite, for the model {\rm MWsat\_n1\_$\Phi 30$}.  A prominent $m=1$ lopsided pattern is excited in the stellar density distribution of MW-like host after each of the pericentre passages. }
\label{fig:density_maps_allmodels}
\end{figure*}

\begin{figure}
\centering
\resizebox{\linewidth}{!}{\includegraphics{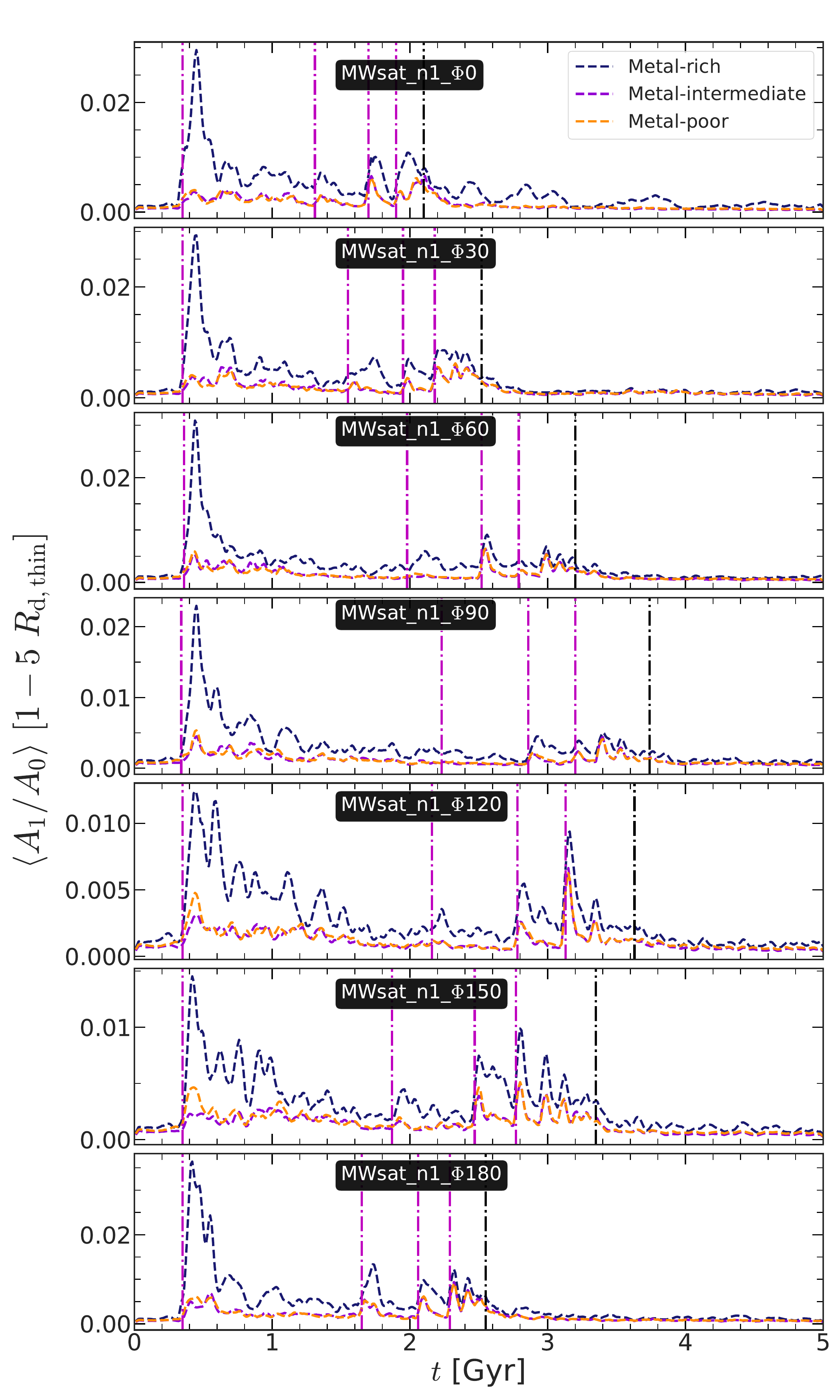}}
\caption{Temporal evolution of the  $m=1$ Fourier amplitude for the stellar density (averaged between $1-5 \ R_{\rm d, thin}$), calculated separately for the metal-rich, metal-intermediate, and metal-poor populations (see the legend) using Eq.~\ref{eq:avg_fourier_density}, for all minor merger models considered here. Vertical magenta dashed-dotted lines denote the epochs of pericentre passages while vertical black dashed-dotted line denotes the epoch of merger. Here, $R_{\rm d, thin} = 4.8 \kpc$. The metal-rich population shows a higher degree of lopsidedness when compared to other two populations.}
\label{fig:lopsided_density_temporal_allmodels}
\end{figure}


Fig.~\ref{fig:density_maps_allmodels} shows the face-on ($x-y$-plane) stellar density distribution (MW-like host and satellite) for metal-rich, metal-intermediate, and metal-poor populations at different times (capturing several pericentre passages and merger of the satellite) for the minor merger model {\rm MWsat\_n1\_$\Phi 30$}. Even a mere visual inspection reveals that the stellar disc of the MW-like host is preferentially perturbed (along an azimuthal direction) each time it experiences a tidal encounter (during the pericentre passage) with the satellite. In other words, the tidal encounter excites a prominent $m=1$ lopsidedness in the stellar disc of the MW-like host. This is a similar dynamical scenario as reported in \citet{Ghoshetal2022}. Furthermore, Fig.~\ref{fig:density_maps_allmodels} suggests that the metal-rich disc of the MW-like host is perturbed to a larger degree when compared to other two populations. By the end of the simulation run ($t = 5 \Gyr$), the satellite has already merged with the MW-like host, and the stellar disc of the post-merger remnant (i.e. MW-like host+satellite) appears to be less perturbed (compare the left-most and right-most panels of Fig.~\ref{fig:density_maps_allmodels}). We checked that the other minor merger models considered here, display a similar dynamical evolution regardless of their varying orbital configurations. 
\par
Before delving into the quantification of properties of the $m=1$ lopsidedness, we mention that after the merger occurs, the stellar and the DM halo particles from the satellite get redistributed in the MW-like host, and the satellite ceases to exist as a separate entity. Therefore, in the post-merger remnant, stellar particles from both the MW-like host and the satellite should be taken into account. Since the satellite contains only 10 percent of the mass of the MW-like host, we checked that the salient findings of this paper (as presented hereafter) remain unchanged even when we considered stellar particles only from the MW-like host. Hence, for all subsequent calculations, we consider stellar particles only for the MW-like host for all times (i.e. before and after the merger phase), unless stated otherwise.
\par 
Next, we quantify the $m=1$ lopsidedness in the stellar disc of the MW-like host as a function of metallicity. As mentioned in sect.~\ref{sec:Intro}, we focus only on the MW-like host while studying the merger driven $m=1$ lopsidedness in the stellar density distribution. To achieve that, we calculated the radial variations ($0-6 \ R_{\rm d, thin}$) of the $m=1$ Fourier coefficients of the mass distribution in the disc of the MW-like host using

\begin{equation}
A_m/A_0 (R)= \left| \frac{\sum_j m_j e^{i m \phi_j}}{\sum_j m_j} \right|\,; \hspace {0.2 cm} \mbox{here,} \  m=1\,.
\label{eq:fourier_calc}
\end{equation}
Here, $A_m$ denotes the coefficient of the $m$th Fourier moment of the density distribution, $m_j$ is the mass of the $j$th particle, and $\phi_j$ is the corresponding azimuthal angle in the cylindrical coordinates \footnote{The summation runs over all the particles within the radial  annulus $[R, R+\Delta R]$, with $\Delta R = 0.25 \kpc$.}. Before characterising the Fourier coefficients, we caution the reader that for the calculation of Fourier coefficients using Eq.~\ref{eq:fourier_calc}, it is extremely crucial to determine the centre and then putting radial annuli centred around that point \citep[for further discussion, see][]{Ghoshetal2022,GhoshandDonghia2025}. In the minor merger models considered here, a tidal encounter with the satellite often induces a tidal tail or the filamentary structures in the outer stellar discs (see in Fig.~\ref{fig:density_maps_allmodels}). In such a dynamical scenario, the centre-of-mass (mass-weighted centre) could be misleading to locate the actual centre of the mass distribution \citep[as shown in][]{Ghoshetal2022}, and in turn, might produce spuriously large $m=1$ amplitude in the central disc region. To avoid that, we first calculated the density-weighted centre of stars using \citep{CasertanoandHut1985}
\begin{equation}
{\bf{x}}_{d,j} = \frac{\sum_{i} {\bf x}_i \rho^{(i)}_j}{\sum_{i} \rho^{(i)}_j}\,,
\label{eq:density_weighted_centre}
\end{equation}
\noindent where ${\bf x}_i$ denotes the three-dimensional position vector for the \textit{i}th particle, and $ \rho^{(i)}_j$ is the density estimator of order \textit{j} around the  \textit{i}th particle, and is evaluated as
\begin{equation}
\rho_j = \frac{j-1}{V(r_j)}m_i\,.
\end{equation}
\noindent Here, $m_i$ denotes the mass of the particle, $r_j$ being the distance of the \textit{j}th particle from the particle around which the local density is estimated. Here, we choose $j=6$, as prescribed by \citet{CasertanoandHut1985}. Throughout the paper, we use the density-weighted centre as the `centre' for the disc of the MW-like host, unless stated otherwise.
\par
In Appendix.~\ref{appen:details_radial_variation_morphokine} (see top panels of Fig.~\ref{fig:radial_fourier_orb30} there), we show the radial profiles of the $m=1$ Fourier coefficients, separately calculated for the metal-rich, metal-intermediate, and metal-poor populations (using Eq.~\ref{eq:fourier_calc}) for the model {\rm MWsat\_n1\_$\Phi 30$}. Initially, before the first pericentre passage, the $A_1/A_0$ values are less than 0.1\footnote{Following \citet{Ghoshetal2022}, throughout this paper, we use $A_1/A_0 = 0.1$ as the onset of an $m=1$ lopsidedness. However, we mention that this is just an operational definition to denote the onset of the lopsidedness.}, thereby indicating the absence of any prominent $m=1$ lopsidedness in the stellar density distribution. However, after each pericentre passage of the satellite, the $A_1/A_0$ values increase well above 0.1, thereby indicating the presence of a prominent $m=1$ spatial or density lopsidedness. This broad trend holds true for all three populations (of different metallicities) considered here, and also for all minor merger models. For the sake of brevity, they are not shown here.
\par
Next, we quantify the temporal evolution of the merger induced $m=1$ lopsidedness in the stellar density distribution of MW-like host as a function of metallicity. We checked that due to the presence of tidal tails and/or filamentary structures (e.g. see in Fig.~\ref{fig:density_maps_allmodels}) in the outer disc region, the corresponding $A_1/A_0$ values often get spuriously large. To avoid such artifact, we restrict our domain of interest to $5 \ R_{\rm d, thin}$ for all minor merger models. \citet{Ghoshetal2022} considered the median value of $A_1/A_0$ (calculated within the radial extent of interest, at time $t$) as the representative value for studying the temporal evolution of the $m=1$ distortion. However, we point out that the usage of the median value of $A_1/A_0$ as the representative one might be affected by occasional fluctuations in the radial variation of $A1/A_0$ values. Instead, in this paper, we follow a more robust method to compute the average (representative) values of $A_1/A_0$ at time $t$. We computed the integrated contribution of the $m=1$ Fourier component over a radial range as \citep{SahaandNaab2013}
\begin{equation}
\avg{A_1/A_0} = \frac{1}{(R_{\rm max}- R_{\rm min})} \int_{R_{\rm min}} ^{R_{\rm max}} A_1/A_0 (R) \ dR\,,
\label{eq:avg_fourier_density}
\end{equation}
\noindent where $R_{\rm max}$ and $R_{\rm min}$ denote the upper and lower boundary points of the integration, respectively. We checked that in our selected minor merger models, the $m=1$ lopsidedness is most prominent within $1-5 \ R_{\rm d, thin}$ ($R_{\rm d, thin} = 4.8 \kpc$). Therefore, for all minor merger models considered here, we set $R_{\rm min} = 1 \ R_{\rm d, thin}$ and $R_{\rm max} = 5 \ R_{\rm d, thin}$. The corresponding temporal evolution of $\avg{A_1/A_0}$, separately calculated for the metal-rich, metal-intermediate, and metal-poor populations (using Eq.~\ref{eq:avg_fourier_density}) are shown in Fig.~\ref{fig:lopsided_density_temporal_allmodels} for all 7 minor merger models considered here. As seen clearly from Fig.~\ref{fig:lopsided_density_temporal_allmodels}, the values of $\avg{A_1/A_0}$ remains close to zero at the beginning of the simulation till the first pericentre passage occurs, thereby indicating the absence of any prominent $m=1$ lopsidedness in the stellar density distribution of the MW-like host. Shortly after the first pericentre occurs, the $\avg{A_1/A_0}$ values display a rapid increase, indicating the onset of a strong $m=1$ lopsided distortion in stellar density. After the pericentre passage, when the satellite goes farther away from the MW-like host, and the MW-like host gets time to readjust itself, the $\avg{A_1/A_0}$ values show a steady decrease, thereby indicating the gradual weakening of the $m=1$ lopsided distortion. Subsequent pericentre passages of the satellite further excites a strong $m=1$ lopsidedness (as indicated by the rapid increase of $\avg{A_1/A_0}$ values). Ultimately, after the satellite merges with the MW-like host, and the post-merger remnant gets enough time to readjust itself, the $\avg{A_1/A_0}$ values show a steady decrease (and close to become zero), implying the $m=1$ lopsidedness fades away progressively. While this broad trend in temporal evolution holds true for all three populations (with different metallicities), however, we find that the $\avg{A_1/A_0}$ values for the metal-rich population always remain much higher when compared with metal-intermediate and metal-poor populations, and this remains true for all times when the system harbours an $m=1$ lopsidedness. In other words, the metal-rich population responds much more vigorously to the perturbing potential, thereby giving rise to a much stronger $m=1$ lopsided distortion. The overall trend of the temporal evolution of the $m=1$ lopsidedness and its associated variation with metallicity of stellar particles remain true for all minor merger models, regardless of their varying orbital configurations (compare different panels in Fig.~\ref{fig:lopsided_density_temporal_allmodels}). We will return to this trend in sect.~\ref{sec:met_kine_connection}. We mention that, the scale length of the metal rich disc ($R_{\rm d, thin}$) is larger than the scale lengths of intermediate and thick discs (for details, see sect.~\ref{sec:equilibrium_model}). Since, in eq.~\ref{eq:avg_fourier_density}, the radius $R$ is normalised by $R_{\rm d, thin}$, therefore, one might wonder whether a larger value of $\avg{A_1/A_0}$ for the metal rich component is truly an intrinsic effect or just because of the fact that the metal rich (or, kinematically coldest) component, most exposed to tidal perturbations, being the most extended. To address this, we checked the radial profiles $\avg{A_1/A_0}$, separately calculated for the metal-rich, metal-intermediate, and metal-poor populations (without any normalisation applied to radius $R$). We find that, indeed the values of $A_1/A_0 (R)$ are larger for the metal-rich component when compared to the other two populations (as also can be partially judged from Fig.~\ref{fig:radial_fourier_orb30}). This trend holds true for all radial ranges where there is a prominent $m=1$ lopsidedness.

\section{Kinematic lopsidedness and its variation with metallicity}
\label{sec:chemical_dissection_Kinelopsided}

\begin{figure*}
\centering
\resizebox{\linewidth}{!}{\includegraphics{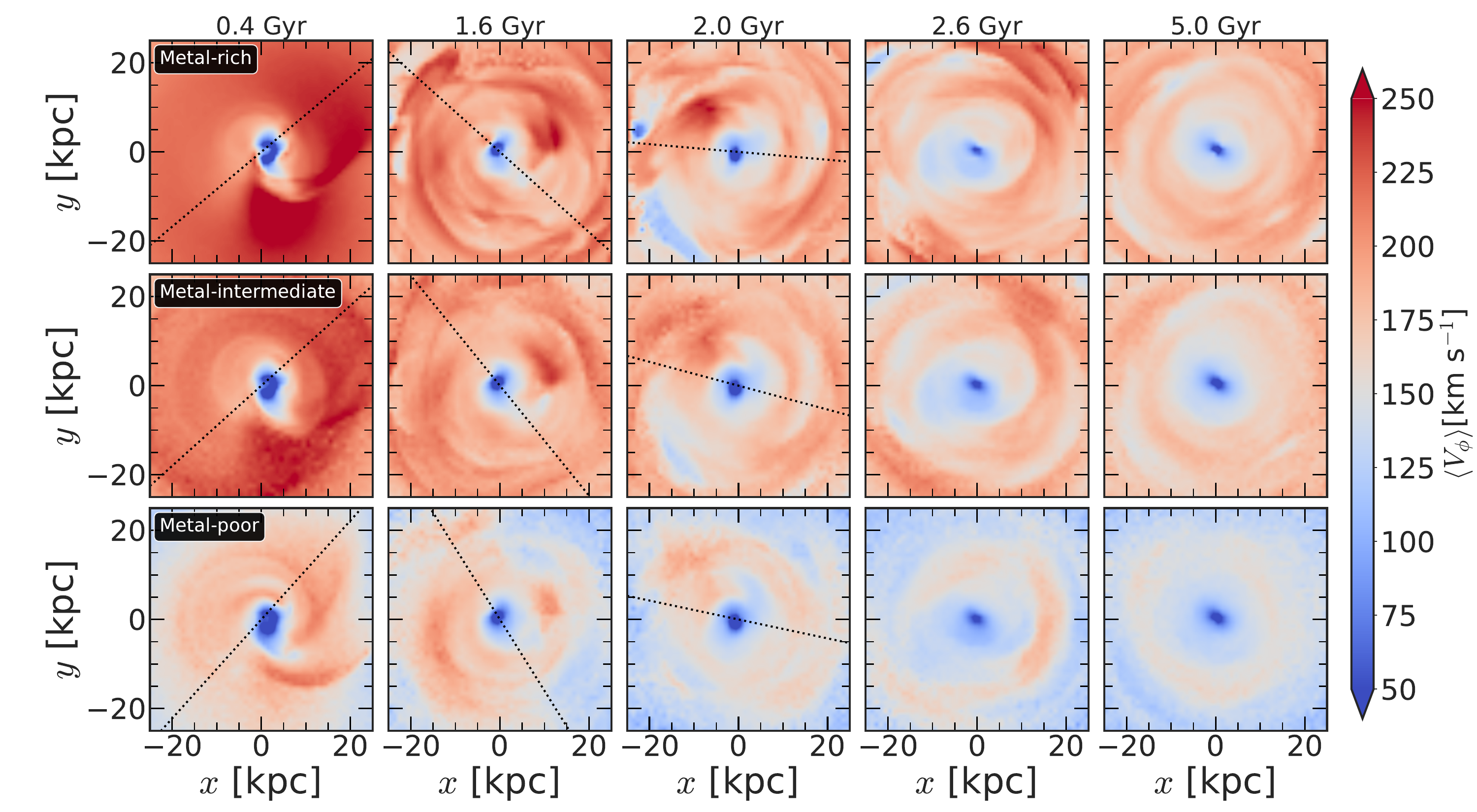}}
\caption{\textit{Chemical dissection of the $m=1$ kinematic lopsidedness:}  Face-on ($x-y$-plane) distribution of the mean azimuthal velocity, \avg{V_{\phi}} (MW-like host only), for the metal-rich (\textit{top row}), metal-intermediate (\textit{middle row}), and metal-poor  (\textit{bottom row}) populations, at different times capturing different pericentre passages and the merger epoch of the satellite, for the model {\rm MWsat\_n1\_$\Phi 30$}. The black dotted lines denote the orientation of the kinematic lopsidedness, i.e., the $\varphi_1 (\avg{V_{\phi}})$ values (using Eq.~\ref{eq:fourier_calc_kineLop}) averaged within $1-5 \ R_{\rm d,thin}$. 
Each of the pericentre passages excites an asymmetry in the \avg{V_{\phi}} distribution.}
\label{fig:vp_maps_allmodels}
\end{figure*}

\begin{figure}
\centering
\resizebox{\linewidth}{!}{\includegraphics{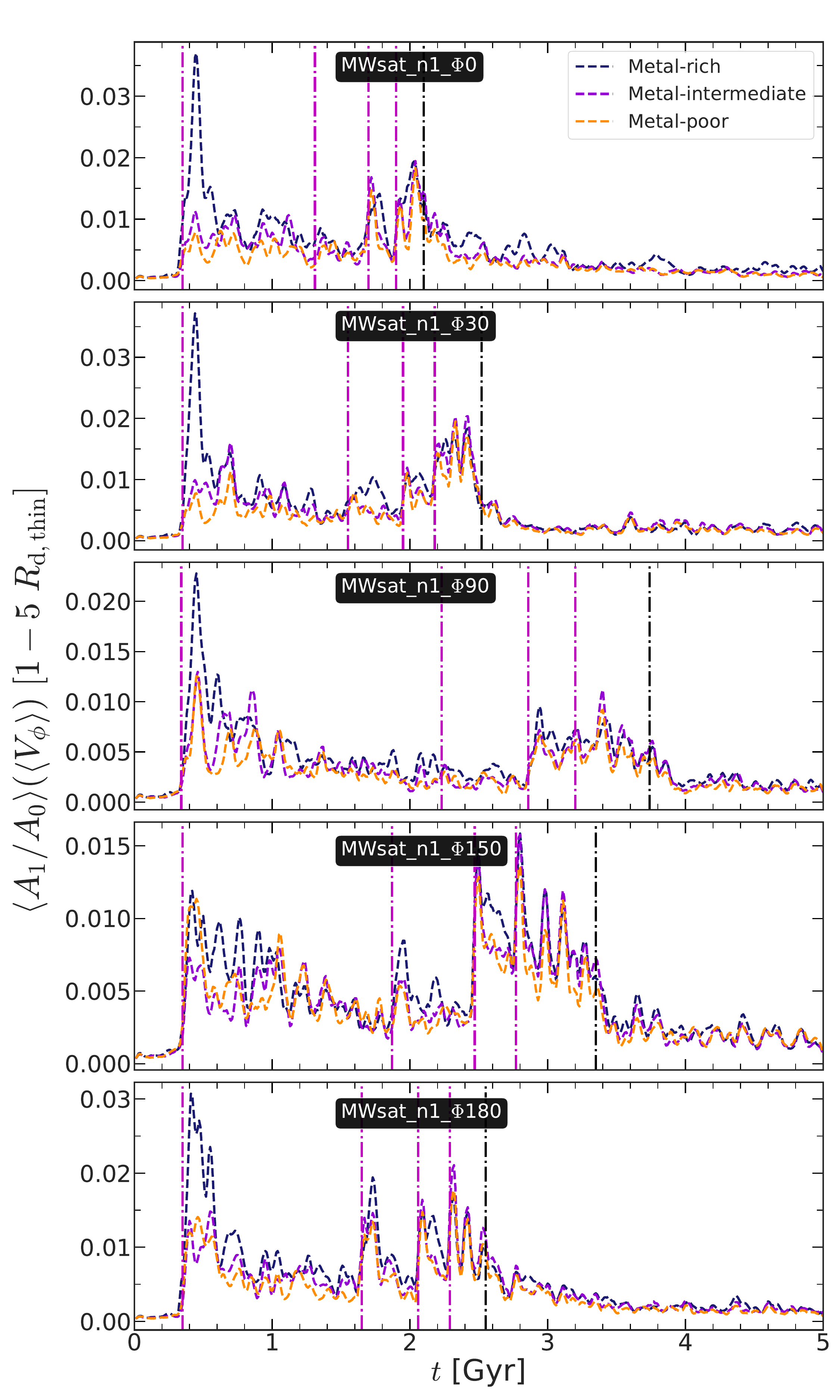}}
\caption{Temporal evolution of the  $m=1$ Fourier amplitude for the stellar mean azimuthal velocity $\avg{V_{\phi}}$ (averaged between $1-5 \ R_{\rm d, thin}$), calculated separately for the metal-rich, metal-intermediate, and metal-poor populations (see the legend), for 5 minor merger models considered here. Vertical magenta dashed-dotted lines denote the epochs of pericentre passages while vertical black dashed-dotted line denotes the epoch of merger. Here, $R_{\rm d, thin} = 4.8 \kpc$. The metal-rich populations shows a higher degree of kinematic lopsidedness when compared to other two populations. }
\label{fig:lopsided_kinematics_temporal_allmodels}
\end{figure}


\begin{figure*}
\centering
\resizebox{0.9\linewidth}{!}{\includegraphics{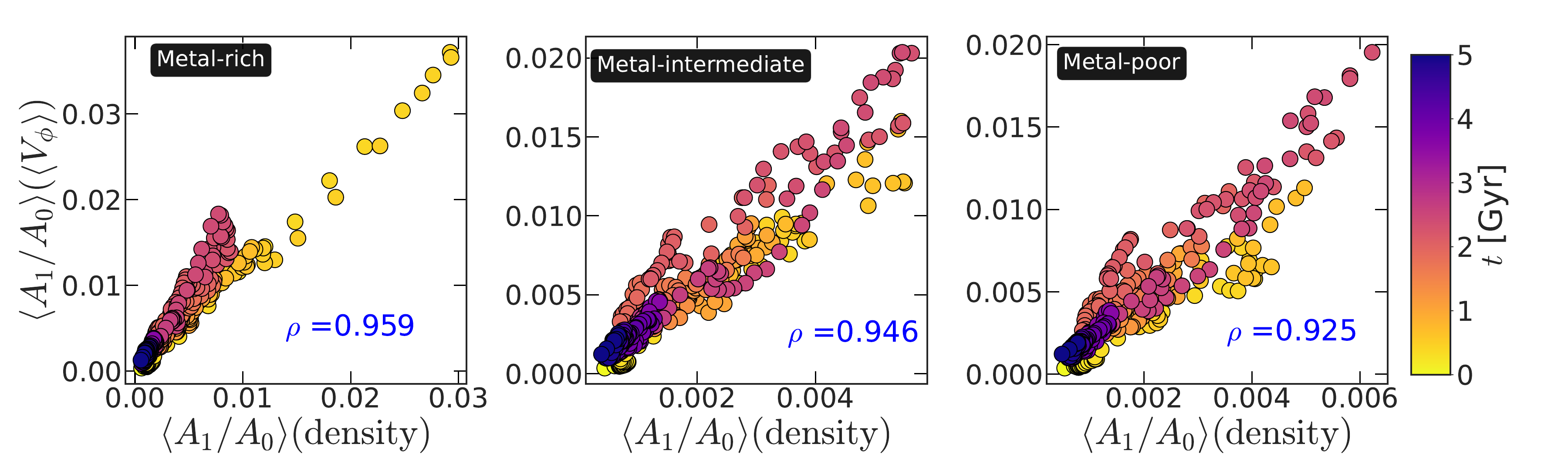}}
\caption{Correlation between the average $m=1$ lopsidedness (computed from the stellar density) and the average $m=1$ kinematic lopsidedness (computed from the stellar $\avg{V_{\phi}}$ distribution), calculated separately for the metal-rich (\textit{left panel}), metal-intermediate (\textit{middle panel}), and metal-poor (\textit{right panel}) as a function of time (see the colour bar), for the minor merger model {\rm MWsat\_n1\_$\Phi 30$}. The corresponding Pearson correlation coefficient, $\rho$ is also indicated in each panel. }
\label{fig:dens_kine_lopsided_CrossCorr}
\end{figure*}

As mentioned in sect.~\ref{sec:Intro}, an $m=1$ lopsidedness in the stellar density distribution would invariably be accompanied by a distortion in the stellar velocity field (i.e. kinematic lopsidedness). This has been demonstrated from semi-analytic theoretical calculations \citep{Jog1997,Jog2002} as well as from observations \citep[e.g. see][]{vanEymerenetal2011} and from numerical simulations \citep{Ghoshetal2022}. Observationally, an off-set between the rotation curves, calculated for the receding
and approaching sides separately for a galaxy, is treated as the signature of a kinematic lopsidedness \citep[e.g. see][]{Swatersetal1999,vanEymerenetal2011}. However, numerical simulation offers a unique opportunity to access the full 6-D phase-space information of stellar particles; thereby facilitating to identify  other measurable signatures of the kinematic lopsidedness. To elaborate, using $N$-body+SPH (smoothed particle hydrodynamics) simulations, \citet{Ghoshetal2022} demonstrated that an asymmetry in the intrinsic mean azimuthal velocity ($\avg{V_{\phi}}$) distribution also bears the signature of an $m=1$ kinematic lopsidedness. Here, we follow the same strategy of using the $\avg{V_{\phi}}$ distribution to study the kinematic lopsidedness in our minor merger models.
\par
Fig.~\ref{fig:vp_maps_allmodels} shows the face-on ($x-y$-plane) distribution of the $\avg{V_{\phi}}$, calculated separately for the metal-rich, metal-intermediate, and metal-poor populations of the MW-like host, at different times for the model {\rm MWsat\_n1\_$\Phi 30$}. A visual inspection of Fig.~\ref{fig:vp_maps_allmodels} reveals the presence of kinematic lopsidedness, that is, if one extracts an one-dimensional velocity profile from this two-dimensional map, along the direction shown by the straight line (in black), the resulting velocity profile would be asymmetric between positive and negative parts of $x$ values \citep[for further discussion, see][]{Ghoshetal2022}. However, we pursue here a more robust method to quantify the kinematic lopsidedness from the $\avg{V_{\phi}}$ distribution of stars. 
\par
The strength of the $m=1$ kinematic lopsidedness is  computed from the $m=1$ Fourier coefficients of the underlying $\avg{V_{\phi}}$ distribution using
\begin{equation}
A_1/A_0 (\avg{V_{\phi}})= \frac{\sum_j m_j V_{\phi,j} e^{i\phi_j}}{\sum_j m_j V_{\phi,j}}\,,
\label{eq:fourier_calc_kineLop}
\end{equation}
\noindent where $V_{\phi,j}$ denotes the value of the azimuthal velocity of the $j$th particle\footnote{We used the same radial binning, as used in Eq.~\ref{eq:fourier_calc}.}. In Appendix.~\ref{appen:details_radial_variation_morphokine} (see bottom panels of Fig.~\ref{fig:radial_fourier_orb30} there), we show the radial profiles of the $m=1$ Fourier coefficients of the $\avg{V_{\phi}}$ distribution, separately calculated for the metal-rich, metal-intermediate, and metal-poor populations (using Eq.~\ref{eq:fourier_calc_kineLop}) for the model {\rm MWsat\_n1\_$\Phi 30$}. The radial profiles of $A_1/A_0 (\avg{V_{\phi}})$ at different times show broadly similar trend as seen for the $A_1/A_0$ profiles computed from the stellar density (using Eq.~\ref{eq:fourier_calc}). To elaborate, the values of $A_1/A_0 (\avg{V_{\phi}})$ remain below 0.1 before the MW-like host experiences any pericentre passage of the satellite (i.e. no prominent $m=1$ kinematic lopsidedness). However, the values of $A_1/A_0 (\avg{V_{\phi}})$ becomes much greater than 0.1 shortly after each pericentre passage (denoting the presence of a strong $m=1$ kinematic lopsidedness).
\par
Next, we quantify the temporal evolution of the $m=1$ kinematic lopsidedness for our minor merger models. Following the same strategy as taken for the $m=1$ lopsidedness in the density field, we calculated the average value of $A_1/A_0 (\avg{V_{\phi}})$,  $\avg{A_1/A_0} (\avg{V_{\phi}})$, within the radial extent $1-5 \ R_{\rm d, thin}$ using Eq.~\ref{eq:avg_fourier_density}. The corresponding temporal evolution of $\avg{A_1/A_0} (\avg{V_{\phi}})$, calculated separately for the metal-rich, metal-intermediate, and metal-poor populations are shown in Fig.~\ref{fig:lopsided_kinematics_temporal_allmodels} for 5 minor merger models considered here. As seen from Fig.~\ref{fig:lopsided_kinematics_temporal_allmodels}, the temporal evolution of  $\avg{A_1/A_0} (\avg{V_{\phi}})$ broadly follows the trend seen for the temporal evolution of $\avg{A_1/A_0}$ calculated for the stellar density field. To explain, at initial times, when the MW-like host has not suffered any pericentre passage of the satellite, the resulting values of $\avg{A_1/A_0} (\avg{V_{\phi}})$ remain close to zero, thereby indicating the absence of a prominent kinematic lopsidedness. However, each pericentre passage is followed by a rapid increase in $\avg{A_1/A_0} (\avg{V_{\phi}})$ values, indicating the presence of a strong $m=1$ kinematic lopsidedness in the MW-like host. After the satellite mergers with the MW-like host and the post-merger remnant gets time to readjust itself, the $\avg{A_1/A_0} (\avg{V_{\phi}})$ values progressively decrease, indicating that the asymmetry in the $\avg{V_{\phi}}$ field ceases to persist. This broad trend is seen for all 7 minor merger models considered here. In addition, the $\avg{A_1/A_0} (\avg{V_{\phi}})$ values for the metal-rich population is larger than those for the other two populations (especially after the pericentre passages), thereby demonstrating that kinematic lopsidedness is much stronger in metal-rich population when compared to other two populations. We will return to this trend in sect.~\ref{sec:met_kine_connection}.
\par
Lastly, we investigate the correlation between the $m=1$ lopsidedness in the stellar density field and the stellar velocity field (i.e. kinematic lopsidedness). To achieve that, we first choose the model {\rm MWsat\_n1\_$\Phi 30$}, and then study the correlation between the $\avg{A_1/A_0}$ and $\avg{A_1/A_0} (\avg{V_{\phi}})$ values, separately for the three chemically-distinct populations, as a function of time. This is shown in Fig.~\ref{fig:dens_kine_lopsided_CrossCorr}. In addition, we computed the Pearson correlation coefficient, $\rho$, for each of these cases. These are also shown Fig.~\ref{fig:dens_kine_lopsided_CrossCorr}. As seen from Fig.~\ref{fig:dens_kine_lopsided_CrossCorr}, the $\avg{A_1/A_0}$ and $\avg{A_1/A_0} (\avg{V_{\phi}})$, denoting the $m=1$ lopsided distortion in the stellar density and the kinematic lopsidedness, show remarkably strong correlation ($\rho > 0.9$) for all three chemically-distinct populations. This demonstrates that the lopsidedness in the density and velocity field evolves in tandem, regardless of their metallicity distribution. We checked the aforementioned correlation for other minor merger models as well, and found the same trend of strong correlations between lopsidedness in the density and the kinematic lopsidedness. For the sake of brevity, they are not shown here. The strong positive correlation between the measured lopsidedness in density and velocity fields  is not surprising as the kinematic lopsidedness can be thought as the representation of the $m=1$ density perturbation, but seen in velocity space.

\section{Disc-DM halo offset and their linkage with merger}
\label{sec:disk_halo_offset}

There has been evidence or indication from past studies of mass modelling from the rotation curve that the baryonic disc and the DM halo may not be concentric (as assumed typically in studies of galactic dynamics), rather they are off-set by some finite amount ($\sim 1- 2.5 \  \kpc$). This includes studies by \citet{Battagliaetal2006} for NGC 5055, for M 99 \citep{Cheminetal2016}, and one galaxy residing in the galaxy cluster Abell 3827 \citep{Masseyetal2015}. In addition, \citet{Kuhlenetal2013}
showed the presence of an long-lived off-set (of $\sim 300 - 400 \pc$) between the density peaks of the baryonic disc and the DM halo in a MW-like galaxy from a high-resolution cosmological hydrodynamical simulation. Furthermore, a recent study by \citet{Ghoshetal2022} showed that tidal interaction with a satellite can be a plausible generating mechanism to excite a disc-DM halo off-set in the host galaxy. However, in their models, the off-set is shown to be short-lived, and is most prominent after each
pericentre passage of the satellite. Studying the generation of such an off-set between the baryonic disc and the DM halo is of great importance in galactic dynamics as an off-centred nucleus can result in an unsettled central region \citep[e.g. see][]{MillerandSmith1992} or a sloshing pattern, seen in a sample of remnants of advanced
mergers of galaxies \citep[e.g. see][]{JogandMaybhate2006}.
\par
\begin{figure}
\centering
\resizebox{\linewidth}{!}{\includegraphics{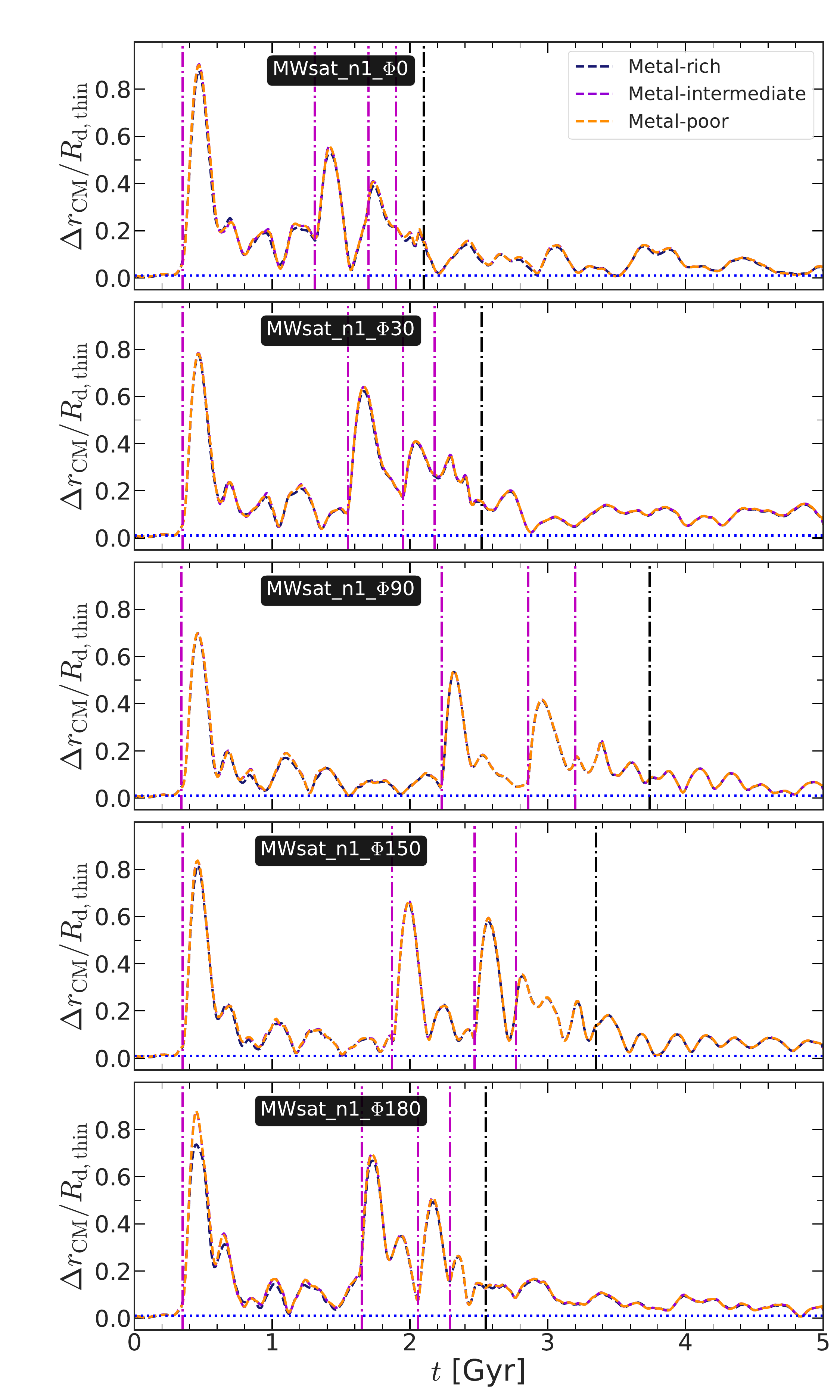}}
\caption{\textit {Disc-DM halo off-set:} Temporal evolution of the off-set between the disc and the DM halo, separately calculated for the metal-rich, metal-intermediate, and metal-poor populations (see the legend) for 5 such minor merger models (with varying orbital configuration). Vertical magenta dashed-dotted lines denote the epochs of pericentre passages while vertical black dashed-dotted line denotes the epoch of merger. The horizontal dotted blue line denotes the softening length of the simulation ($\epsilon = 50 \pc$). Here, $R_{\rm d, thin} = 4.8 \kpc$. Each pericentre passage excites a prominent disc-halo off-set, regardless of the orbital configuration.}
\label{fig:disc_halo_offset}
\end{figure}

Here, we investigate whether a tidal interaction (during a pericentre passage) with a satellite excites an off-set between the baryonic disc and the DM halo in the MW-like host galaxy. Furthermore, we investigate whether such an off-set shows any dependence on the chemical distribution of the stellar population. To achieve that, we first calculated the density-weighted centre of the baryonic disc of the MW-like host (separately for metal-rich, -intermediate, and -poor populations) and the DM halo while using Eq.~\ref{eq:density_weighted_centre}. Next, at time $t$, we computed the separation between the baryonic disc and the DM halo, $\Delta r_{\rm CM}$ using 
\begin{equation}
\Delta r_{i, \rm CM} (t) =\sqrt{\Delta x_{i, \rm CM}^2 (t)+\Delta y_{i, \rm CM}^2 (t)+\Delta z_{i, \rm CM}^2 (t)}\,,
\label{eq:disc_dm_offset}
\end{equation} 
\noindent where, $\Delta x_{i, \rm CM} (t) = x_{d,i}(t) -x_{\rm dm} (t)$, $\Delta y_{i, \rm CM} (t) = y_{d, i}(t) -y_{\rm dm} (t)$, $\Delta z_{i, \rm CM} (t) = z_{d,i}(t) -z_{\rm dm} (t)$. Here, $i$ denotes metal-rich, metal-intermediate, and metal-poor populations, and  ($x_{d,i} (t), y_{d,i} (t), z_{d,i} (t)$) and ($x_{\rm dm} (t), y_{\rm dm} (t), z_{\rm dm} (t)$) denote the density-weighted centres of the baryonic disc and the DM halo of the MW-like host at time $t$, respectively. Fig.~\ref{fig:disc_halo_offset} shows the corresponding temporal evolution of $\Delta r_{\rm CM}$, calculated for three different populations (with varying metallicities) in 5 such minor merger models considered here. A visual inspection of Fig.~\ref{fig:disc_halo_offset} clearly reveals the excitation of a prominent disc-DM halo off-set in the MW-like host as a dynamical impact of the tidal interaction with the satellite. The off-set can be as high as one disc scale length ($R_{\rm d, thin} = 4.8 \kpc$). The off-set is transient (short-lived) and is most prominent shortly after the pericentre passage occurs, similar to past findings \citep[e.g.][]{Pardyetal2016,Ghoshetal2022}. After the satellite merges with the MW-like host, the off-set $\Delta r_{\rm CM}$ decreases steadily, and by the end of the simulation run ($t = 5 \Gyr$), the value of $\Delta r_{\rm CM}$ approach to zero, thereby indicating that the baryonic disc and the DM halo has again become concentric. However, we do not find any characteristic variation in the temporal evolution of $\Delta r_{\rm CM}$ as a function of metallicity distribution of the stellar disc (i.e. metal-rich, metal-intermediate, and metal-poor), and this trend holds true for all minor merger models considered here. The short-lived nature of the disc-halo off-set could potentially explain the (apparent) dearth of off-centred disc and DM halo configuration, as reported in cosmological simulations \citep[e.g. see][]{GaoandWhite2006}.

\section{Connection between metallicity and kinematics: Effect on lopsidedness}
\label{sec:met_kine_connection}

\begin{figure}
\centering
\resizebox{0.88\linewidth}{!}{\includegraphics{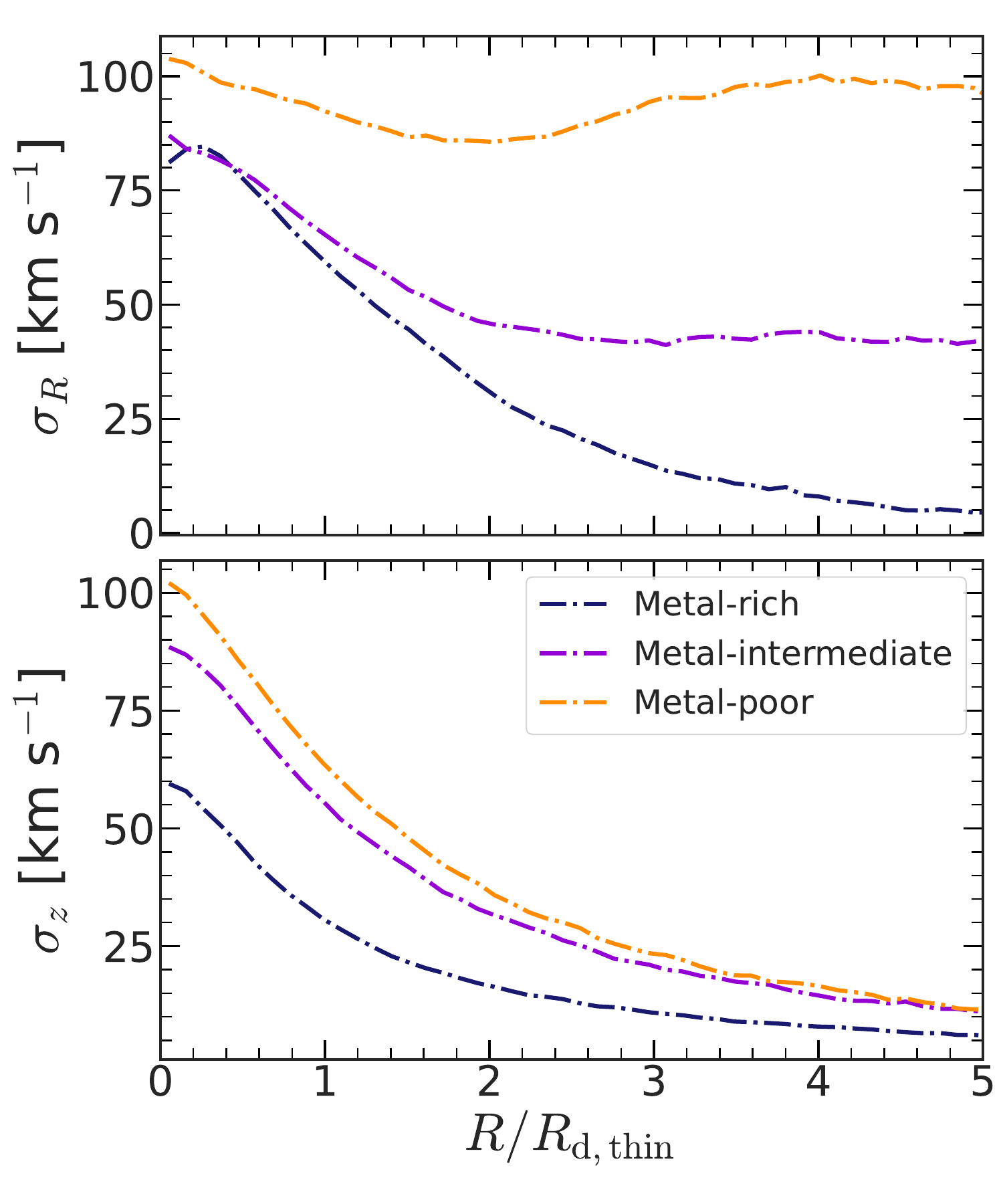}}
\caption{Radial profiles of the radial (\textit{top panel}) and vertical (\textit{bottom panel}) velocity dispersion, for the metal-rich, metal-intermediate, and metal-poor populations of the MW-like host (see the legend), calculated at $t =0$ for the model {\rm MWsat\_n1\_$\Phi 30$}. Here, $R_{\rm d, thin} = 4.8 \kpc$. Metal-rich population is kinematically cold when compared to other two populations.}
\label{fig:kine_chem_connection}
\end{figure}

In previous sections, we demonstrated that the $m=1$ Fourier coefficients, calculated both from the stellar density and velocity ($\avg{V_{\phi}}$) fields, are much higher in the metal-rich population of the MW-like host when compared to other two populations (i.e. metal-intermediate and metal-poor). In other words, the $m=1$ or lopsided feature is stronger in the stellar density as well as the kinematics of the metal-rich population. In addition, we find that this dependence of the amplitude of the $m=1$ lopsidedness on the metallicity of stellar particles holds true for all seven minor merger models considered here, regardless of their orbital parameters. 
Next,  we investigate in detail to physically understand this trend. First, we mention that the metal-intermediate and the metal-poor populations, by construction, are vertically thicker (i.e. higher vertical scale heights) and more centrally concentrated (i.e. shorter disc scale length). Furthermore, in Fig.~\ref{fig:kine_chem_connection}, we show the radial variations of the radial and vertical velocity dispersions, $\sigma_R$ and $\sigma_z$, as a function of metallicity, calculated at $t=0$ for the model  {\rm MWsat\_n1\_$\Phi 30$}. By construction  of the equilibrium models, both, the $\sigma_R$ and $\sigma_z$ values for the metal-poor population, are set to be much higher \citep[for details, see][]{Pagninetal2023} when compared to those for the metal-rich populations (with the metal-intermediate population falling in between). In other words, the metal-rich population is kinematically much colder when compared to the other two populations. During the course of several pericentre passages and the subsequent mergers, the MW-like host does get dynamically heated up (i.e. it shows an increase in velocity dispersion), however, we checked that the velocity dispersion for the metal-rich population always remains lower than that for the other two populations. We argue that since the metal-rich population is kinematically cold, it responds to the external perturbing potential more vigorously than the other two populations, thereby harbouring a much stronger $m=1$ lopsidedness in stellar density and velocity fields when compared to the metal-intermediate and metal-poor populations. We checked that this trend remains true for all seven minor merger models considered here. This trend is analogous to the stronger disc response shown by low velocity dispersion stars and gas, to an $m=2$ potential compared to that shown by high velocity dispersion stars, as seen in the case of $m=2$ spiral arms \citep[e.g. see][]{Rohlfs1977} , and for $m=2$ bars \citep[]{Ghosetal2023,Ghoshetal2024}. The findings presented here, reinforces the fact that a higher response of a lower velocity dispersion component to an external perturbation is a generic feature.

\section{Variation of m=1 lopsidedness with metallicity in the LMC}
\label{sec:implications_LMC}
In previous sections, we systematically showed that a metal-rich and kinematically cold (i.e. lower velocity dispersion) disc is more susceptible to external perturbing source (i.e. tidal encounter with the satellite), and hence displays a stronger $m=1$ lopsidedness in the stellar density as well as velocity fields of an MW-like host. This trend is shown to hold true for a wide variation of orbital parameters.
\par
\begin{figure}
\centering
\resizebox{\linewidth}{!}{\includegraphics{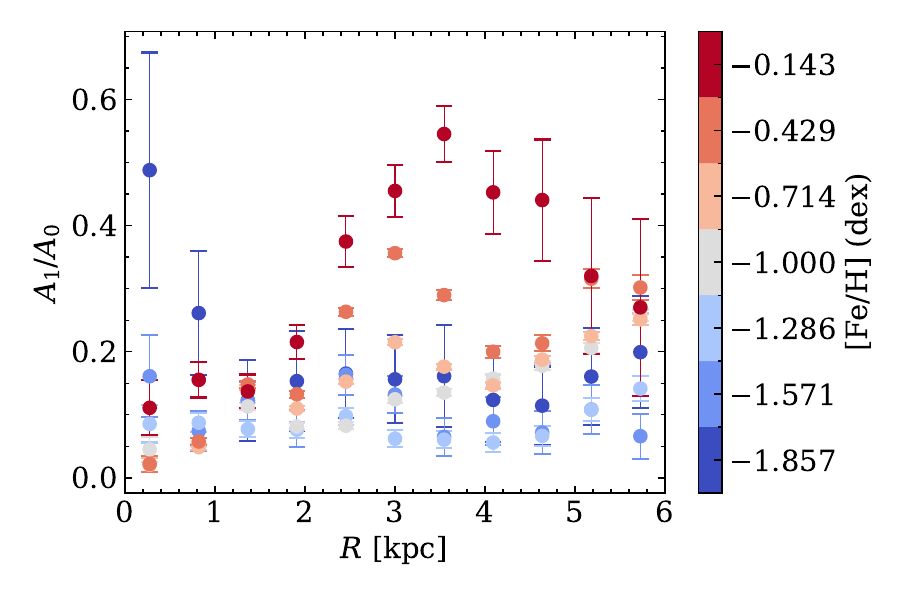}}
\caption{Radial profiles of the $m=1$ Fourier moment, computed from the stellar density distribution of the LMC. Using the $\feh$ estimates from \citet{Frankeletal2025} based on the method outdlined in \citet{Andrae_rix_chandra_2023}, the stars are sub-divided into 7 metallicity bins. Metal-rich stellar population shows a higher degree of $m=1$ lopsided distortion (i.e. larger $A_1/A_0$ values) at almost all radial locations. The uncertainties represent the standard deviation of the best-fit estimates for $A_1/A_0$ over 100 bootstrap samples from the dataset. }
\label{fig:lopsidedness_in_LMC}
\end{figure}

 
Here, using \gaia\ DR3 phase-space information and the metallicity estimates from the XP spectra and photometry \citep{Andrae_rix_chandra_2023, Frankeletal2025}, we investigate whether the $m=1$ structure in the stellar density distribution of the LMC displays any characteristic variation with the chemical composition of stars.  For the sample of stars of the LMC, selected from the \gaia\ DR3 stars, associated quality cuts, and estimations of metallicity, the reader is referred to \citet{Frankeletal2025}. First, we sub-divide
the stars into same 7 metallicity bins, as done in \citet{Frankeletal2025}, and compute
the corresponding radial variation of the $m=1$ Fourier amplitude of the stellar density of the LMC. This is shown in Fig.~\ref{fig:lopsidedness_in_LMC}. We bootstrap stars from the sample 100 times to estimate uncertainties on the Fourier parameters. As seen clearly, indeed the metal-rich stars tend to show a larger degree of $m=1$ distortion (i.e. higher value of $A_1/A_0$), and the most metal-poor population shows the least prominent $m=1$ distortion (i.e. lower value of $A_1/A_0$). However, we mention that the $m=1$ component with a radially varying phase, which appears like a one-armed spiral, has a stronger amplitude (especially in the inner region of the LMC).
It is now observationally well known that the LMC harbours an $m=1$ spirals \citep[e.g. see][]{Ruiz-Laraetal2020,Jimenez-Arranz2022}. 
In literature, both an one-arm spiral and an $m=1$ lopsidedness are considered as an $m=1$ distortion (i.e. yielding an non-zero value of $A_1/A_0$), and the main difference lies in the radial variation of the corresponding phase-angle $\varphi_1$ \citep[for a detailed discussion, see][]{JogandCombes2009}. In fact, interactions with a satellite can trigger both an an $m=1$ spirals and an $m=1$ lopsidedness in stellar disk of the host, as shown in \citet{Ghoshetal2022}.

%
\par 
From the trend shown in Fig.~\ref{fig:lopsidedness_in_LMC}, it is tempting to draw the conclusion that for the LMC, a dynamical scenario similar to what is shown for the MW-host from our minor merger models, holds true as well.  
However, we point a few subtle differences between our models and the LMC. \citet{Frankeletal2025} showed that for stars with $\feh < -0.1$ dex, the scale length of the exponential disc diminishes progressively with increasing $\feh$  values which is broadly opposite to how we modelled the host and satellite in our minor merger models. Secondly, whether the metal-rich stars in the LMC are indeed kinematically-cold - it remains to be verified. Furthermore, the LMC is influenced by the SMC (Small Magellanic Cloud) and also possibly by the MW \citep[for a detailed discussion, see][]{Vasiliev2023} - a dynamical scenario which is essentially different from our MW-host+satellite set-up. However, on general dynamical grounds, it can be argued that in a tidal interaction with a massive host galaxy, a satellite of lower mass undergoes a stronger tidal disturbance, including a disruption and merger in some cases. The larger galaxy, on the other hand, undergoes a smaller distortion including lopsidedness, as in the MW-LMC interaction \citep [see, e.g,] []{Weinberg1995}. This point is confirmed by the result that the asymmetry in the LMC is shown to arise due to interaction with the smaller mass SMC (Small Magellanic Cloud), rather than by an interaction with the MW \citep[e.g.,][]{DiazandBekki2012, Beslaetal2012,Beslaetal2016}.
 Further, it has long been known that the SMC is subdivided or split into two main components \citep [e.g.,][]{Mathewsonetal1986}; where the interaction between the LMC and the SMC that occurred about $\sim 2 \times 10^8$ yr ago has been shown to lead to a break-up of the SMC, and also result in an asymmetry in the LMC \citep[e.g.,] []{MuraiandFujimoto1980,DiazandBekki2012,Beslaetal2012}. This indicates that to examine the variation of the strength of lopsidedness as a function of metallicity 
in the LMC, as shown in Fig.~\ref{fig:lopsidedness_in_LMC}, an interaction with a smaller mass satellite galaxy would need to be studied, where the LMC is the larger-mass galaxy in the interacting pair. This will be pursued in a future work.
\par 

To conclude, while the broad physical understanding for a metallicity-dependent variation of the $m=1$ distortion is laid out in this work, and it is shown that in the LMC, the degree of $m=1$ distortion shows an increasing trend with stellar metallicity (Fig.~\ref{fig:lopsidedness_in_LMC}), however, it remains to be examined the connection between metallicity of stars and their associated velocity dispersion (to be treated as a measure of kinematic coldness).

\section{Discussion}
\label{sec:discussion}

Here, we mention some implications and a few limitations of this work. First, the minor merger models we used are dissipationless, high-resolution $N$-body simulations, that is, the models lack any gas dynamics, star formation recipe and chemical enrichment model. This restricts us from studying the properties of $m=1$ lopsidedness in gas as well as investigating any correlation between the $m=1$ lopsidedness seen in stars and gas. In addition, in the scheme of assigning the metallicity (a posteriori) to the stellar particles, as done here, does not include any radial and vertical metallicity gradient. However, such metallicity gradient exists in the MW \citep[e.g. see][]{Bovyetal2012,Haywoodetal2013,Bensbyetal2014,Haydenetal2015}. Although the findings in this work clearly show a variation of $m=1$ lopsidedness with metallicity and kinematic properties of a stellar population, it will be worth checking the properties of the $m=1$ lopsidedness when gas dynamics and chemical enrichment model are incorporated in models in a self-consistent fashion. 
\par
Secondly, due to the tidal interaction, the MW-like host also harbours an $m=2$ bar (and associated) spiral in the central disc region. Although a tidal origin of a bar is well known \citep[e.g. see][]{Thompson1981,Giuricinetal1993,Anderson1996,Barazzaetal2009,M_ndez_Abreu_2012,Linetal2014,Ghoshetal2021}, it would be worth investigating how the temporal evolution of the $m=2$ bar and an $m=1$ lopsidedness are connected, as a function of chemical distribution of stars. Furthermore, Fig.~\ref{fig:density_maps_allmodels} suggests that the $m=2$ bar might be off-set wrt the baryonic disc. A systematic study of the plausible off-set bar and its connection with the $m=1$ lopsidedness, as a function of metallicity and for varied orbital configuration will be pursued in a future work.

\section{Summary}
\label{sec:conclusion}

 In summary, we carry out a systematic study of the variation of the $m=1$ lopsidedness in stellar density and velocity distribution as a function of chemical distribution of stars. To achieve that, we make use of a total of 7 dissipationless, high-resolution $N$-body simulations of minor merger models with a MW-like host and a satellite (of mass ratio 1:10) with varying orbital configuration. For the chemical distribution, we assigned, a posteriori, metallicities to the stellar particles of the MW-like host and the satellite based on their kinematic properties
and the current observational constraints. We follow the excitation of an $m=1$ lopsidedness due to a tidal encounter, and the subsequent temporal evolution in the MW-like host as a function of metallicity. Our main findings are listed below.

\begin{itemize}

\item{The tidal encounter (during the pericentre passage) excites a prominent $m=1$ lopsidedness in the stellar density distribution of the MW-like host. After the merger happens, the strength of $m=1$ lopsidedness decays steadily, and in extreme cases, it ceases to exist. The metal-rich population always shows a larger degree of lopsidedness when compared to metal-intermediate and metal-poor populations. We found that these trends are generic across all minor merger models (with varying orbital configuration) considered here.}

\item{The tidal encounter with the satellite also excites a prominent $m=1$ kinematic lopsidedness in the MW-like host. The temporal evolution of the $m=1$ asymmetry in the stellar velocity field broadly follows the evolution of the $m=1$ lopsided distortion in the density, and the corresponding strengths are seen to be strongly correlated (i.e. Pearson correlation coefficient, $\rho > 0.9$). Furthermore, the strength of the $m=1$ kinematic lopsidedness show a similar metallicity dependence as seen for the $m=1$ distortion in the density, with metal-rich population showing the strongest lopsidedness in stellar velocity.}

\item{We mention that in our minor merger models, the metal-rich stars are kinematically-colder (i.e. lower velocity dispersion) when compared to metal-intermediate and metal-poor populations; thereby making them more susceptible to external perturbations. We propose that this could be the underlying physical reason that explains the stronger $m=1$ lopsidedness, seen in both, the stellar density and velocity fields, of the MW-like host, when compared to other two populations.}

\item{Using the \gaia\ phase-space information, together with the metallicity estimates from \citet{Andrae_rix_chandra_2023, Frankeletal2025}, we show that in the LMC, the strength of the $m=1$ distortion in stars increases with metallicity. The non-zero values of $A_1/A_0$, the $m=1$ Fourier amplitude, are predominantly due to an $m=1$ spiral arm (where the phase varies with radius), especially in the inner stellar disc of the LMC. }

\item{The tidal encounter also excites an off-set between the baryonic disc and the DM halo of the MW-like host. The off-set is seen to be short-lived, and is most prominent immediately after the pericentre passages of the satellite. This trend holds true for all minor merger models (with different orbital configuration) considered here. However, we do not find any significant variation in the disc-halo off-set with metallicity of stars in the MW-like host.}

\end{itemize}

 To conclude, the lower velocity dispersion of the metal-rich (thin) disc makes it more susceptible to the formation of lopsidedness in it. The velocity dispersion of a disc component can be taken to be a dynamical proxy to explain the metallicity dependence of the strength of lopsidedness, as shown in this work.

\section*{Acknowledgments}

S.G. thanks Hans-Walter Rix for helpful discussions at the initial phase of this work. S.G. acknowledges funding from the IIT-Indore, through a Young Faculty Research Seed Grant (project: `INSIGHT'; IITI/YFRSG/2024-25/Phase-VII/02). PDM acknowledges the support of the French
Agence Nationale de la Recherche (ANR), under grant ANR-13-BS01-0005
(project ANR-20-CE31-0004-01 MWDisc). C.J.J. acknowledges support from INSA, New Delhi as an INSA Senior Scientist. This work has made use of the computational resources obtained through the DARI grant A0120410154 (P.I. : P. Di Matteo).


\bibliographystyle{mnras}
\bibliography{my_ref.bib} 

\begin{appendix}

\section{Merger history of the minor merger models}
\label{appen:details_merger_history}

\begin{figure}
\centering
\resizebox{0.75\linewidth}{!}{\includegraphics{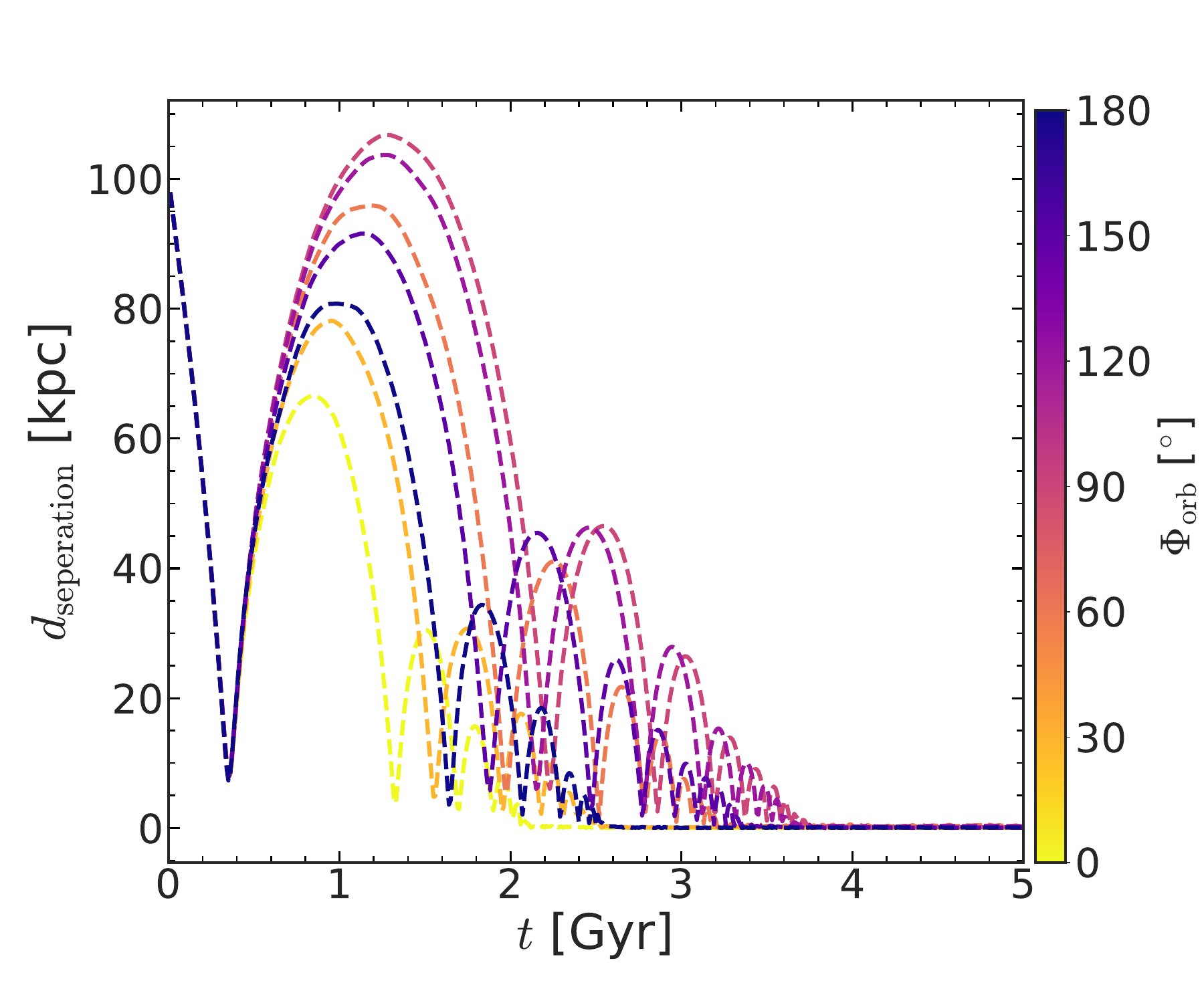}}
\caption{Temporal variation of the separation between the MW-like host and the satellite, $d_{\rm separation}$, for all 7 minor merger models with different orbital angles ($\Phi_{\rm orb}$, see the colour bar). Different models show different epochs of merging (i.e. $d_{\rm separation} \rightarrow 0$), depending on the orbital configuration. for details, see the text.}
\label{fig:MW_Sat_distance_allmodel}
\end{figure}

Here, we briefly mention some of the parameters related to merger history of minor merger models considered here. First, we calculated the separation between the MW-like host and the satellite, $d_{\rm separation}$, by computing the distance between the density-weighted centres (using Eq.~\ref{eq:density_weighted_centre}) of the MW-like host and the satellite. The temporal evolution of $d_{\rm separation}$ for all 7 minor merger models (with varying orbital configuration) is shown in Fig.~\ref{fig:MW_Sat_distance_allmodel}. In each case, the satellite has gone through a series of pericentre passages. During each pericentre passage, the satellite is subjected to a dynamical friction, and loses a finite amount of its orbital angular momentum, resulting in progressively smaller $d_{\rm separation}$ at the time of  pericentre (see Fig.~\ref{fig:MW_Sat_distance_allmodel}). Table.~\ref{table:key_param_mergers} provides the epochs of (first) four pericentre passages for each minor merger model. Furthermore, we define the epoch of merger, $t_{\rm merger}$ as the time when $d_{\rm separation}$ approaches to zero. The corresponding epochs of mergers for all 7 models are also given in Table.~\ref{table:key_param_mergers}. For a given orbital configuration (prograde or retrograde), the epoch of merger progressively increases with increasing values of $\Phi_{\rm orb}$ (see table.~\ref{table:key_param_mergers}).

%
\begin{table*}
\centering
\caption{Key parameters of the merger history of minor merger models.}
\begin{tabular}{ccccccc}
\hline
\hline
Model  & $t_{\rm peri, 1}$ (Gyr) & $t_{\rm peri, 2}$ (Gyr) & $t_{\rm peri, 3}$ (Gyr) & $t_{\rm peri, 4}$ (Gyr) & $t_{\rm merger}$ (Gyr) & $t_{\rm end}$ (Gyr)\\
\hline
{\rm MWsat\_n1\_$\Phi 0$} & 0.35 & 1.31 & 1.7 & 1.9 & 2.1 & 5\\
{\rm MWsat\_n1\_$\Phi 30$} & 0.35 & 1.55 & 1.95 & 2.18 & 2.52 & 5  \\
{\rm MWsat\_n1\_$\Phi 60$} & 0.36 & 1.98 & 2.52 & 2.79 & 3.2 & 5\\
{\rm MWsat\_n1\_$\Phi 90$} & 0.34 & 2.23 & 2.86 & 3.2 & 3.74 & 5  \\
{\rm MWsat\_n1\_$\Phi 120$} & 0.35 & 2.16 & 2.78 & 3.13 & 3.63 & 5  \\
{\rm MWsat\_n1\_$\Phi 150$} & 0.35 & 1.87 &  2.47 & 2.77 & 3.5 & 5  \\
{\rm MWsat\_n1\_$\Phi 180$} & 0.35 & 1.65 & 2.06 & 2.29 & 2.55 & 5  \\
\hline
\end{tabular}
\newline{
\textbf{Notes:} From left to right : name of the minor merger model, epochs of different pericentre passages, epoch of merger, and the total time over which the simulation is evolved.}
\label{table:key_param_mergers}
\end{table*}

\section{radial variation of the m=1 morpho-kinematic lopsidedness}
\label{appen:details_radial_variation_morphokine}

\begin{figure*}
\centering
\resizebox{0.9\linewidth}{!}{\includegraphics{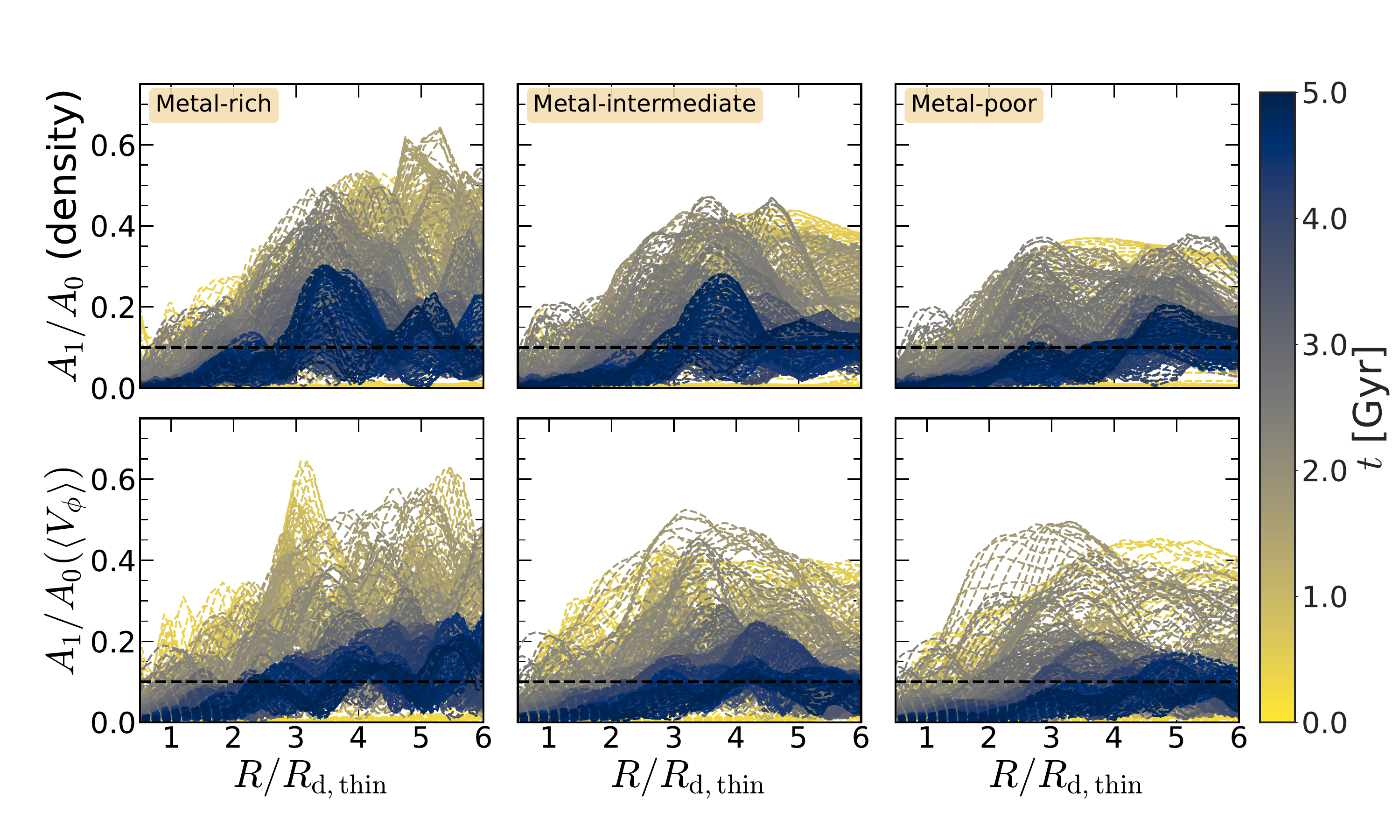}}
\caption{Radial variation of the $m=1$ Fourier moment, computed from the stellar (MW-like host only) density (\textit{top row}) and from the mean azimuthal velocity, $\avg{V_{\phi}}$ (\textit{bottom row}) as a function of time (see the colour bar) for the model {\rm MWsat\_n1\_$\Phi 30$}.  \textit{Left panels} show for the metal-rich population while \textit{middle panels} and \textit{right panels} show for the metal-intermediate and metal-poor populations, respectively. Here, $R_{\rm d, thin} = 4.8 \kpc$. The horizontal black dashed line denotes the threshold value of $A_1/A_0 = 0.1$. For further deatils, see the text.}
\label{fig:radial_fourier_orb30}
\end{figure*}

Fig.~\ref{fig:radial_fourier_orb30} (\textit{top panels}) shows the radial variation of the $m=1$ Fourier coefficients at different times, computed from the stellar density distribution of the MW-like host, for the metal-rich, metal-intermediate, and metal-poor populations of the  model {\rm MWsat\_n1\_$\Phi 30$}. 
Fig.~\ref{fig:radial_fourier_orb30} (\textit{bottom panels}) shows the radial variation of the $m=1$ Fourier coefficients at different times, computed from the $\avg{V_{\phi}}$ distribution, for the metal-rich, metal-intermediate, and metal-poor populations of the minor merger model {\rm MWsat\_n1\_$\Phi 30$}. Each pericentre passage excites a prominent $m=1$ asymmetry in both the stellar density and velocity field (kinematic lopsidedness) of the MW-like host.

\end{appendix}


\label{lastpage}
\end{document}